\documentclass[journal]{IEEEtran}
\usepackage{xcolor,soul,framed} 
\usepackage{graphicx}
\usepackage{amsmath}
\usepackage{array}
\usepackage{mdwmath}
\usepackage{mdwtab}
\usepackage{eqparbox}
\usepackage{url}
\usepackage{lineno,hyperref}
\usepackage{amsmath,amssymb}
\usepackage{booktabs}
\usepackage{makecell}
\usepackage{multirow}
\usepackage{graphicx}
\usepackage{multirow}
\usepackage{tabu,bm}
\usepackage{color}
\usepackage{subfigure}
\usepackage{lipsum,mwe,cuted}
\usepackage{stfloats}
\usepackage{cite}
\usepackage{enumitem} 
\usepackage{diagbox} 

\begin{document}

\title{Nine Challenges in Artificial Intelligence and Wireless Communications for 6G}

\author{
Wen Tong\thanks{Dr. Wen Tong (tongwen@huawei.com) is with Wireless Technology Labs, Huawei Technologies, Ottawa, Canada.} and
Geoffrey Ye Li \thanks{Prof. Geoffrey Ye Li (Geoffrey.Li@imperial.ac.uk) is with Intelligent Transmission and Processing (ITP) Lab., Dept. Electrical and Electronic Engineering, Imperial College London, United Kingdom.\\The main content of this article was presented at talks from the authors.} 
}

\maketitle
\begin{abstract}

In recent years, techniques developed in {\it artificial intelligence} (AI), especially those in {\it machine learning} (ML), have been successfully applied in various areas, leading to a widespread belief that AI will collectively play an important role in future wireless communications. To accomplish the aspiration, we present nine challenges to be addressed by the interdisciplinary areas of AI/ML and wireless communications, with  particular focus towards the {\it sixth generation} (6G) wireless networks. Specifically, this article classifies the nine challenges into computation in AI, distributed neural networks and learning, and ML enabled semantic communications.

\end{abstract}

\begin{IEEEkeywords}
computing crisis, distributed learning, and semantic communications, 6G wireless networks. 
\end{IEEEkeywords}

\section{Introduction}

By late 1980s, the basic ensemble of algorithms for computational neural networks has been nearly completed even though the size of the neural networks nowadays is much larger. Back to 30 years ago, a neural network under exploratory study typically consists of only a handful of neurons configured over a few layers. It can only  handle a limited amount of training data due to the constraints of computing resources and complexity, which limits its capability, performance, and  applications. A few scholars persisted through early 2000s. They were able to produce new thinking, which resulted in the huge success of {\it deep learning} (DL) and led them wining  the prestigious Turing Award in 2018. 

Computing power plays a critical role in the development of {\em artificial intelligence} (AI).  Following the Moors’ Law, the computing power  has been increased by around 200,000 times compared with 30 years ago, which has improved the capability of machine learning (ML) by 100,000 times. For a deep neural network, each layer consists of millions or even tens of millions of neurons with hundreds of millions of parameters. Thanks to the efforts of those few dedicated scholars, brute force computation, and big data, DL has surpassed the human level in many applications. 

By the way, when we were PhD students in electrical engineering and computer science around 30 years ago, we took some courses on neural networks. However, the seniors also told us it was hard to find jobs if studying ML. Therefore, we shifted our interest into wireless communications and participated in and led the investigation from the first generation (1G) to the fifth (5G) generation wireless networks.

\begin{figure*}[!t]
\centerline{\includegraphics[width=5in]{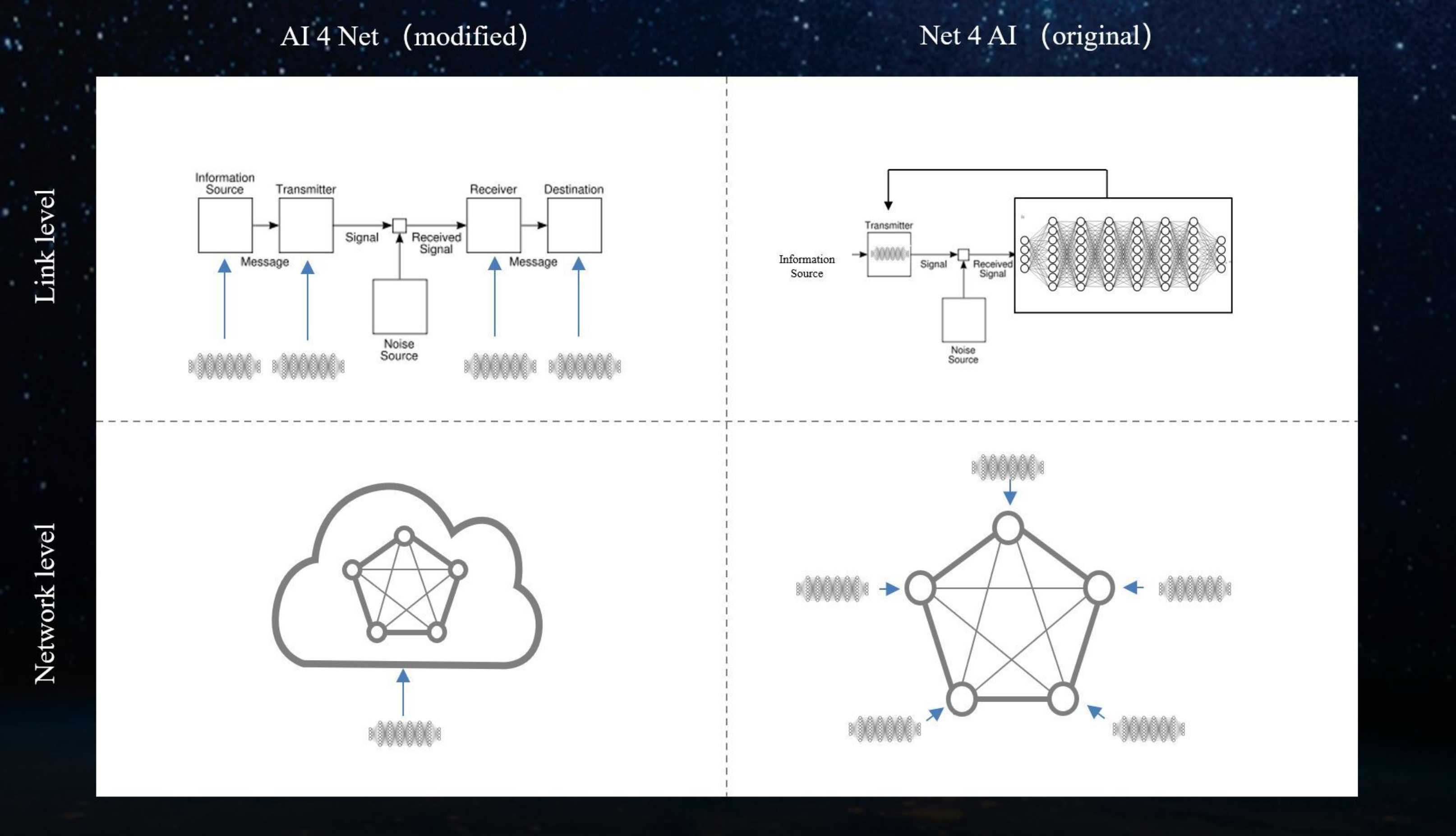}}
\caption{Comparison of AI4Net or Net4AI at link level and network level, respectively.}
\label{fig1}
\end{figure*}

In the next 10 to 20 years, our general mission is to facilitate various applications of ML technologies through intelligent communications. We would like to clarify a couple of concepts between AI and wireless networks, including {\it AI for communication network} (AI4NeT) and {\it communication network for AI} (Net4AI).  In Figure \ref{fig1}, we demonstrate their correlation and difference at the link level and the network level, respectively. Particularly, AI4Net mainly belongs to 5G and 5.5G while Net4AI is the mainstay for the sixth generation (6G) and future wireless networks. Nowadays, we are applying ML techniques to improve the intelligence and transmission performance of communication networks. For 6G and beyond, AI will be used everywhere in our daily life and communication networks will  transmit and collect massive amount data required by ML. Therefore, AI will revolutionize 6G and will also be the fundamental business for 6G. AI in 6G is far more than just {\em over-the-top} (OTT) applications. For 6G and beyond, we would put particular focus on Net4AI.

The brute-force computing will be a stimulant for the development of AI for a while. The new understanding, new discoveries, and new innovations are largely dependent on the computing power of AI. Moreover, the success and popularization of AI  require endless big data, which relays on future wireless networks for data transmission and collection. Unfortunately, neither brute-force computing nor endless big data is sustainable. Therefore, we identify nine challenges that should be addressed for 6G, which are very critical and to be detailed in the following. If there is no theoretical breakthrough in the related areas, we still have to grope in the dark. Hopefully, the interleaving of AI and 6G wireless networks will shed a light on producing new theories for future wireless networks.

In this article, we will discuss nine fundamental scientific challenges in communications and AI with focus on the problems and challenges in the interdisciplinary areas of DL and  wireless communications, especially those key challenges for ensuring successful 6G wireless networks. Particularly, they include general challenges in deep neural networks for wireless communications, distributed DL and distributed neural networks, and DL enabled semantic communications.

\section{General Challenges in Deep Neural Networks for Wireless Communications}

In the following, we will discuss five fundamental challenges in ML with close relation to wireless communications for 6G.

\subsection*{Challenge 1: Computing Crisis of DL}

\begin{figure*}[!t]
\centerline{\includegraphics[width=5in]{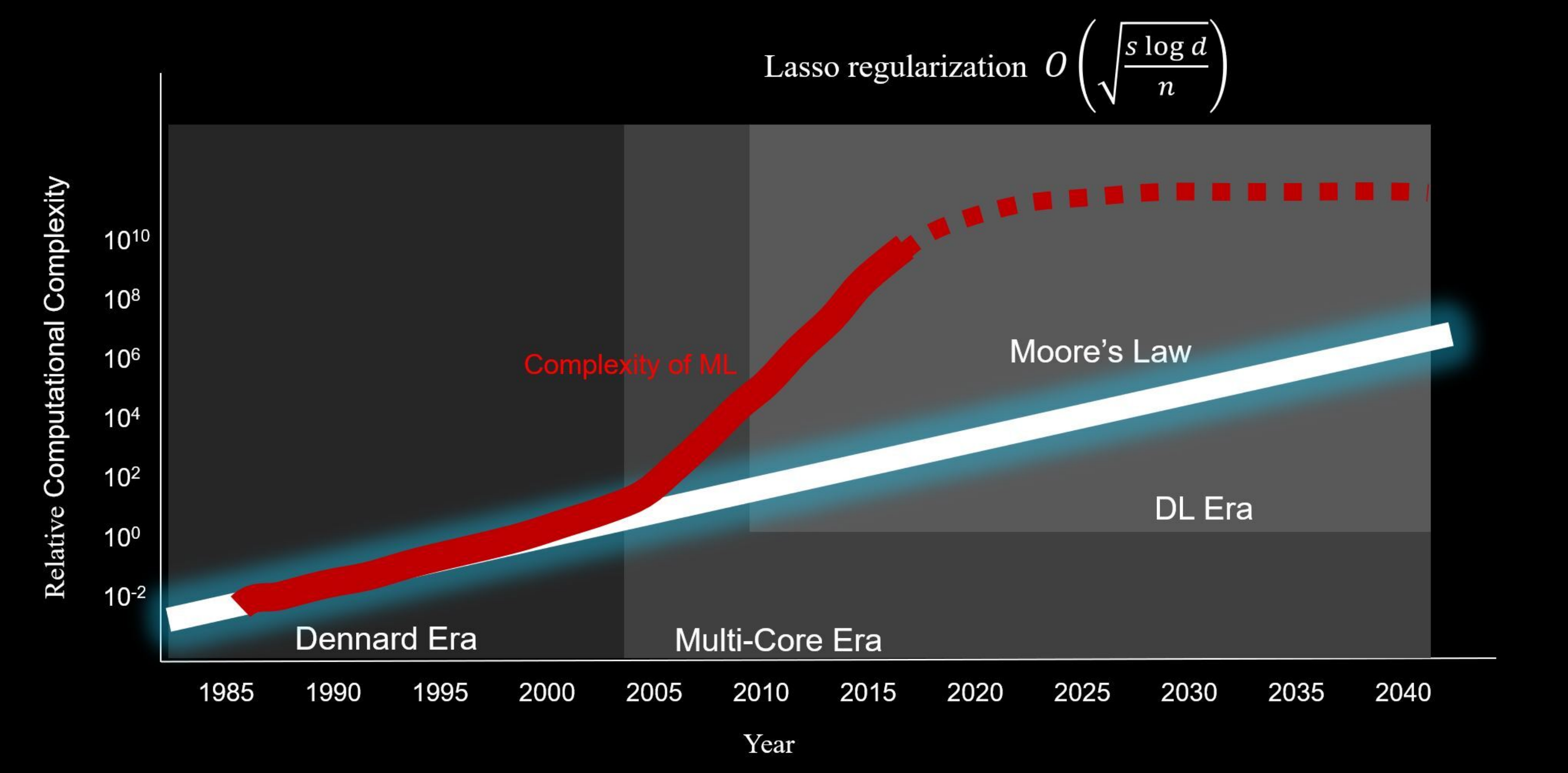}}
\caption{Computing complexity at different eras.}
\label{fig2}
\end{figure*}

Based on the research from the MIT team in \cite{MIT1}, Figure \ref{fig2} compares the complexity of ML with the Moore's Law in different eras. During the Dennard growing era, the computing capability is improved by increasing the clock speed, the power consumption of AI computation almost follows the Moore's Law. It starts to grow faster than the Moore's Law during the multi-core era. Today we have entered the era of DL. Due to the demand on brute-force computing in AI, the power consumption increases far exceeding the growth rate of the Moore's Law, at least with $10^5$. Therefore, it is necessary to take energy consumption into consideration to ensure sustainable development. Otherwise, our efforts in the area may be in vain.

Moreover, to reduce the error rate of ML to as low as 5\%, the computation cost is extremely expensive, which requires 100 billion US dollars and 10 quintillion program instructions~\cite{MIT1}. Therefore, reducing the computational complexity becomes the primary task for the future AI. Even if some  techniques, such as pruning, low-dimensional compression, less quantization levels, or smaller DNNs, can be applied to mitigate the issue, the AI computing crisis is still a long-term engineering problem in the following decades.

To address the above challenges, we should first establish a set of evaluation methodologies for power consumption in computation and communication  to facilitate the selection of  solutions.

\subsection*{Challenge 2: Gradient Disappearance in DL}

The {\em backward-propagation} (back-propagation) gradient iteration in DL is a very headache issue for hardware acceleration architecture, which affects wireless transmission latency and is of great importance for the end-to-end communications\cite{Hao}.

In the past, many neural network  patching methods have been developed to address the issues on the back-propagation of iterative gradient descent algorithms. Some of them are listed as following:
\begin{itemize}
\item Saturation of sigmoid function, such as {\em rectified linear unit} (ReLU) function, causes gradient disappearance. The negative semi-axis of the ReLU function is a dead zone, where the gradient turns into zero. Therefore, {\em leaky version of ReLU} (LeakyReLU) and {\em parametric ReLU} (PReLU) have been proposed to substitute the ReLU. 
\item To strengthen the stability of the gradient and weight distribution, {\em exponential linear unit} (ELU) and {\em scaled ELU} (SELU) are adopted in deep neural networks. 
\item It is difficult to pass gradients if a neural network is too deep. Therefore, the highway network \cite{Q7} has been invented, in which the parameters are even omitted and {\em residual networks} (ResNets) are used instead.
\item {\em Batch normalization} (BatchNorm) can stabilize the mean and the variance of neural network parameters.
\item DropOut, which ignores some units/neurons during the training, can increase noise to the gradient flow, therefore, it can mitigate overfitting and lower the generalization error.  
\item The gradient in {\em recurrent neural networks} (RNN) is sometimes unstable. So {\em long short-term memory} (LSTM) is widely used, which can be further improved by adding some {\em gated recurrent units} (GRUs).
\item The {\em Jensen–Shannon} (JS) divergence in the {\em generative adversarial network} (GAN) causes gradients to disappear or become invalid, which can be solved by the {\em Wasserstein GAN} (WGAN).
\end{itemize}
The above are just some patching methods, but they are unable to fully address the issue .

By re-examining the Arnold-Kolmogorov's arbitrary function approximation theorem, it may be possible to develop a one-step or an iterative algorithm to address the issue on iterative gradient descent in DL.  Through in-depth research on the nature and the behavior of the  back-propagation in DL, perhaps with some theory developments, we may be able to mitigate or  address the issue completely. Of course, this is a long-term goal and we are looking forward to a major breakthrough one day in the near future.

\subsection*{Challenge 3: Memory Capacity of Deep Neural Networks}

It is predicted that the memory function of neural networks will be fully exploited in the future communications. For example, semantic transmitter and receiver require a huge memory capacity to store the background knowledge\cite{Semantic}. In general, the larger the memory capacity of a neural network, the higher potential to improve communication efficiency, especially the efficiency of semantic communications to be introduced in Section IV. A natural question turns out to be how many types of objects or events a neural network can remember or what is the memory capacity of a neural network.

It has been shown in \cite{Hopfield} that the memory capacity of a Hopfield network with $n$ neurons is 
\begin{equation}
    C \approx \frac{n}{2 \log_2 n}. 
\end{equation}
However, the memory capacity is still unknown for a general deep neural network.

Another correlated question is the relationship between the memory capacity and the computational complexity. Is it linear, exponential, or some other relations? Such a relationship is of great importance for inference and estimation at the receiver in  communication systems. 
 
\subsection*{Challenge 4: Dependence of DL on Big Data (I still think is very confusing)}

The big data required by training deep neural networks is often collected by wireless communications. Therefore, it becomes critical how to train neural networks efficiently and quickly and consequently to lower the requirement on big data and wireless communications. 

Even if the relationship between the memory capacity and the size of deep neural networks is still not clear as indicated in Challenge 3, the neural network with a large memory capacity is usually with a massive number of parameters and requires a huge amount of data to train the model. 

There have been some initial works on exploiting the domain knowledge in the area of communications to reduce the requirements on training data, such as model-driven DL for communication systems \cite{Q9}. It is a  tricky issue to trade off the domain knowledge usage, big data requirements, communication performance, and system complexity. It is also important to know the minimum amount of data required for a given communication applications in general. 

\subsection*{Challenge 5: Dynamic, Accretionary, and Meta Learning}

Currently,  AI models are trained under special learning hypothesis premise that the environment (the statistics of the training data) is static, at least during the training period. However, real communication scenarios, especially for mobile communications, are constantly changing. The basic features of dynamic DL for wireless communications require extensive investigations. 

The challenge can be potentially addressed with the developments of accretionary learning and meta learning. The theory of accretionary learning for human beings was proposed by cognitive psychologists over 40 years ago. Recently, the concept of accretionry learning\cite{Fred} has been developed for classification. Particularly, traditional ML trains a classification model with fixed recognition sets, which fails for the data set with new features. The accretionary learning\cite{Fred} can address the issue by extending  deep neural networks through interaction of accretion, fine-tuning, and restructuring. Moreover, meta learning\cite{Meta}, also known as learning-to-learn, becomes an active research area recently. With its development, dynamic learning for communication systems and networks could be achieved one day.  

\section{Distributed Neural Networks and Distributed Learning}

Nowadays, there are around 7 millions base stations and 6 billions mobile terminals in the world and these numbers keep on growing.  It is impossible to collect all data from different base stations and terminals into a super fusion server for centralized processing due to huge storage requirement, heavy computing complexity, demanding energy consumption, and privacy issues. Therefore, distributed neural networks and distributed learning become extremely important in the future wireless communications.

\subsection*{Challenge 6: Wireless Data Aided Collective Learning}

{\em Collective Learning} is referred to that multiple servers collect data and train ML models collectively. It belongs to a general category of distributed learning, but with special relation to communications.   

The architecture and performance of collective learning depend on both AI itself and communications. In the integrated architecture of AI and wireless communications, the primary problems are how to segment data and how to separate communication and computing, where the  data is mainly used for training ML model while the communication model is used for data/parameter transmission. Here, we pay close attention to  communication issues.

Specifically, how to segment the neural network and the tasks for collective learning is an unsolved problem, which is also a profound theoretical issue. Specifically, the related research problems include:
\begin{itemize}
    \item How to collectively learn multiple small and identical neural networks?
    \item How to segment a large network into multiple small networks?
    \item What are the impacts of the size and the number of  networks on the efficiency and optimization of learning?
\end{itemize}

\begin{figure*}[t]
\centerline{\includegraphics[width=5in]{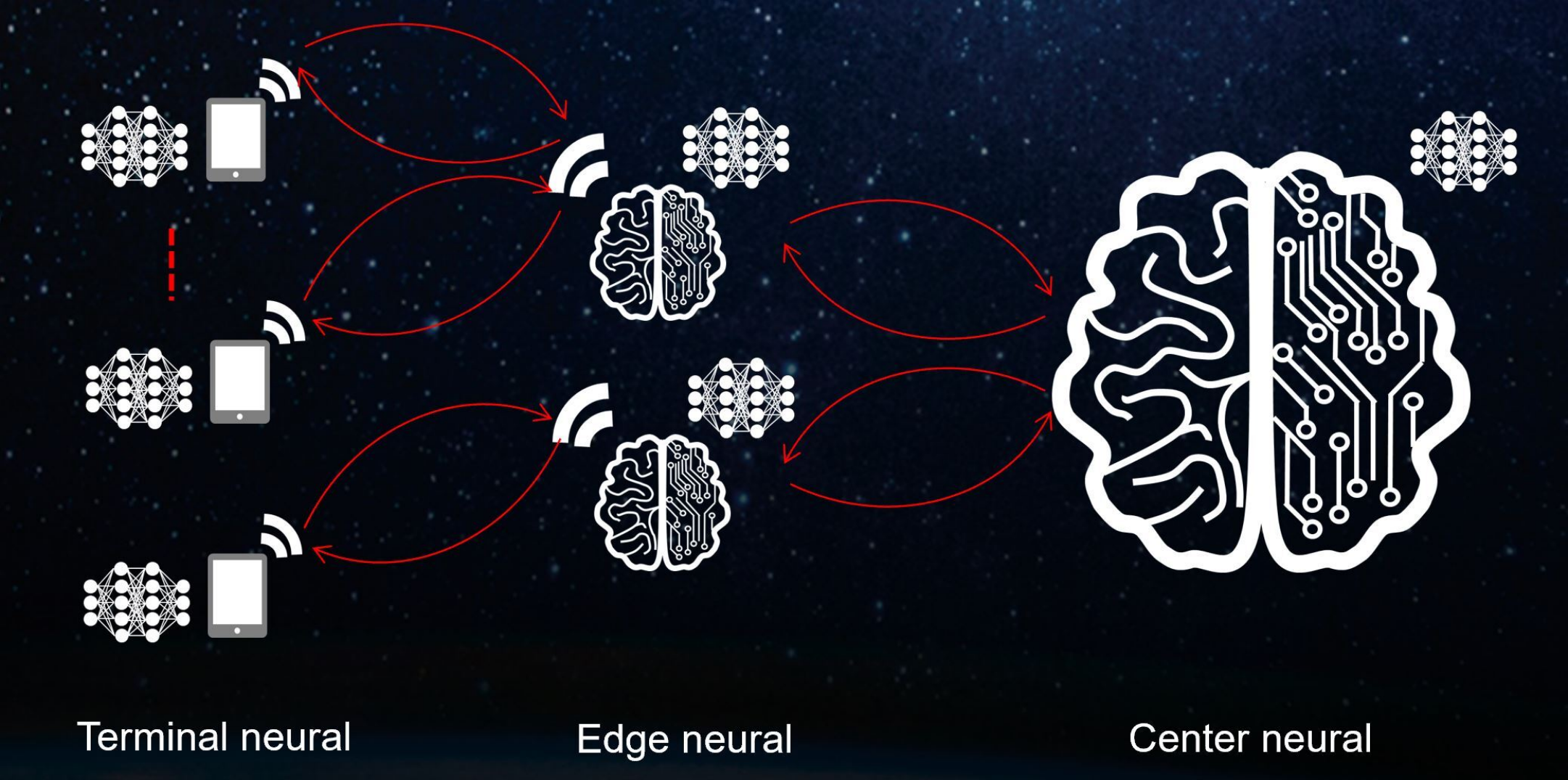}}
\caption{Distributed learning.}
\label{fig3}
\end{figure*}

\subsection*{Challenge 7: Wireless Communications Enabled Federated Learning}

Federated learning \cite{Q3a,Q3b,Q3Qin}, a type of distributed learning, is enabled by wireless communications. Different from the collective learning in Challenge 6, federated learning relies more on communications. As shown in Fig. \ref{fig3}, multiple clients/terminals jointly train/learn the same model in federated learning and there are only parameter exchanges among multiple terminals, rather than segmenting data, tasks, and neural networks as that in the collective learning. The information exchange between the terminal and edge neural networks  depend on wireless communications. The same for that between the edge neural networks and the central networks. Therefore, the transmission error  affects distributed/federated learning significantly.  

In the past, the fields of wireless communications and AI have been developed independently  with different optimization goals. Particularly, communication systems are usually designed to maximize data transmission rate and minimize  bit-error rate while the performance of AI is typically measured by the recognition accuracy and the quality of the reproduced data. In general, the interaction and connection between AI and wireless communications are new trend that has never been jointly investigated before.   

Let us take the learning process of a deep neural network as an example. Even if the impact of the coefficient error in the first few iterations is trivial, it becomes very critical and sensitive in the last few iterations of the last few layers. Therefore, at different iterations and  different layers of a neural network, the demands for the quality of wireless transmission are various. The similar issue shows in inference models.

In addition to the aforementioned error sensitivity issue, there are many others. For example, how to compress model parameters to save communication resources and how to perform robust parameter aggregation at the central server to address communication errors and stragglers, as indicated in \cite{Q3Qin}. These issues pose innovation opportunities for 6G. It is more than desired to find out the relationship between ML model errors and various wireless transmission errors/imperfections from a theoretical basis to facilitate their joint optimization\cite{Hao2}.

\section{Semantic Communications} 

Semantics communications is a problem in the post-Shannon era. Around 70 years ago, Weaver \cite{Shannon} classified communications into three levels in his masterpiece co-authored with Shannon, which include transmission of symbols, semantic exchange, and effect of semantic exchange. When Shannon established information theory, he only focused on the mathematical theory of symbol transmission. Particularly, the amount of information in a source is indicated by its {\em entropy} and the capacity of a communication channel is proved to be the maximum mutual information between the channel input and output, both are measured by bits. Due to the lack of general mathematical tools to model and analyze semantics or meaning of the transmitted contents, limited progress has been made in the past even if many outstanding researchers have tackled the area.

\begin{figure*}[!t]
\centerline{\includegraphics[width=6in]{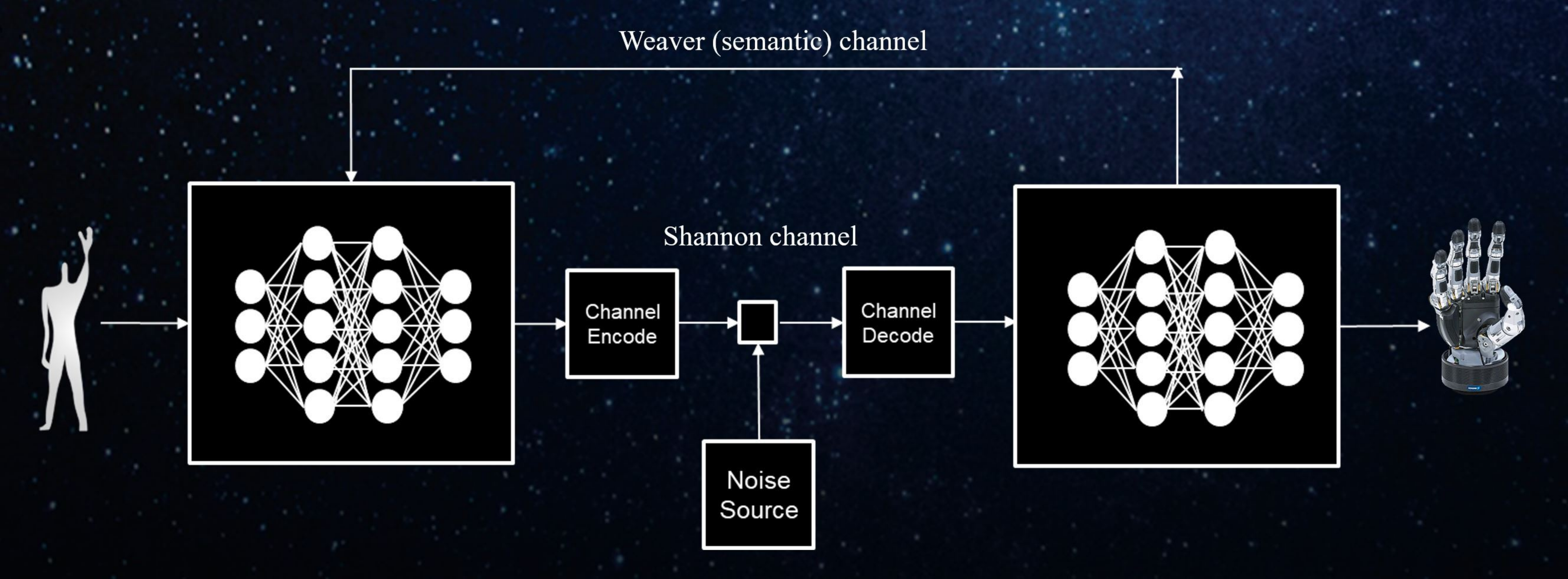}}
\caption{Shannon channels and Weaver channels for typical communications and semantic communications.}
\label{fig4}
\end{figure*}

Recent success in DL and its application makes it possible to re-examine Weaver's second-level and even third-level models, where the goal is not to transmit bit sequence, but the meaning of the content. We then colloquially call them {\em semantic communications}. The preliminary communication architecture is shown  in Figure \ref{fig4}, where the Shannon channel is embedded inside and the Weaver semantic channel is outside. Under this architecture, the scientific questions include: 
\begin{itemize}
\item How do we measure semantics? 
\item What is the limit of semantic communication systems?
\item Is there any theory analogical to Shannon information theory? 
\item How to implement and perform semantic communications efficiently?   
\end{itemize}

It is noticed that semantic communications have caught extensive attentions in the past several years and there have been some initial research in this area\cite{Semantic,Q5, Qin}. We believe that there will be more works on theory and implementation of semantic communications in the near future.

\subsection*{Challenge 8: Mathematical Foundation of Semantic Communications}

Kolmogorov, a great mathematician in the last century, came into contact with Shannon information theory from the other side of the Iron Curtain of the Cold War in the Soviet Union in 1958 and further expanded the mathematical foundation of information theory. Based on his investigation, the sufficient and necessary condition for the mathematical equivalence of a large class of automorphism systems is that they are with the same {\em entropy}. Kolmogorov also extends the mathematical foundation of Shannon entropy into a new interdisciplinary area of topology and algebra. 

The new mathematical foundation for post-Shannon communications or semantic communications is also particularly important. The entropy and capacity in Shannon information theory are defined based on statistical probability, which may not be applicable to semantic communications. 

To develop mathematical foundation of semantic communications, the statistical probability is substituted by logical probability of a message in \cite{Semantic}, where semantic source, semantic noise, semantic entropy, semantic capacity, and semantic coding are defined accordingly. Moreover, semantic source theorem and semantic channel coding theorem, analogical to their counterparts in Shannon information theory, have been proved in \cite{Semantic}. However, it is more than desired to build a comprehensive mathematical framework to identify the optimal structure for semantic communication systems. 

\subsection*{Challenge 9: Structure of Semantic Communication Systems}

Perhaps, many of us have ignored Pages 5 - 7 of the Shannon's masterpiece \cite{Shannon}, which are the earliest statement on natural language AI. Based on Dewey's statistical results on the English language, Shannon attempted to reconstruct English sentences from random English letters by exploiting the probability and statistical methods in order to investigate semantic communications. At that time, Shannon partially retreated to the fundamental probability to support the concept of entropy in information theory, which actually ignored the important elements in semantic communications. 

In \cite{Q5}, the classic Huffman coding with the deep neural network method have been compared for transmission of text in 2018. It has been found that the improvement through using a deep neural network is pretty limited, which again confirmed Shannon information theory. 

It is a critical issue whether the information restoration in semantic communications should be through a faithful recovery of the corresponding bits or direct semantic restoration. It can be regarded as a structural problem in semantic communications as indicated by Shannon and Weave. The question here turns into whether to use a general DL neural network for training and learning for semantic communications or to further explore different structural levels of communications.

\begin{figure}[!t]
\centerline{\vspace{.0in}\hspace{.8in}\includegraphics[width=5.0in]{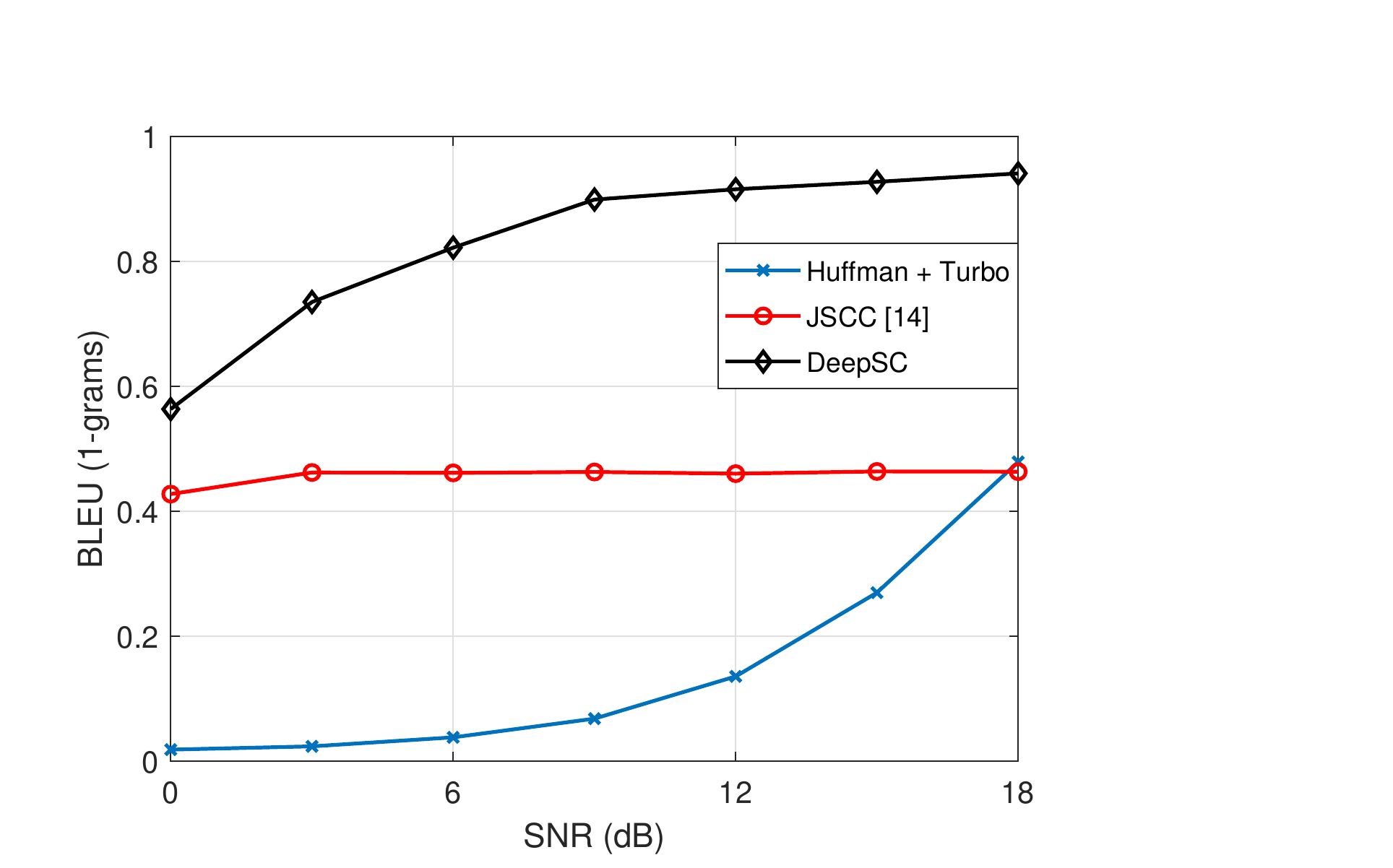}}
\caption{Robustness of the semantic communication system, DeepSC~\cite{Qin}, in comparison to benchmarks.}
\label{fig5}
\end{figure}

Recently, a DL-based semantic communication system for text transmission, named DeepSC, has been developed in \cite{Qin}. With the aid of Transformer and novel loss function design,  DeepSC aims at maximizing the system capacity and minimizing the semantic errors by recovering the meaning of sentences, rather than bit- or symbol-errors as in the traditional communications. To measure the performance of semantic communications properly, a new metric, named sentence similarity, has been proposed. Compared with the traditional communication system without considering semantic information exchange,  DeepSC is more robust to channel variations and is able to achieve better performance, especially in the low signal-to-noise (SNR) regime, as shown in Figure \ref{fig5}. The initial results in \cite{Qin} demonstrate the power of semantic communications. More research in this area, such as semantic communications for video, image, and speech transmission, is expected. 

\section{Conclusion Remarks}

In this article, we discussed nine challenges to ensure successful 6G, including
\begin{itemize}
\item	the computing challenge in deep learning (DL) (Challenge 1)
\item	The communication related challenges in deep DL (Challenges 2 - 5)
\item	The challenges on the architecture of distributed learning enabled by communications (Challenges 6 and 7)
\item	The challenges in DL enabled semantic communications (Challenges 8 and 9)
\end{itemize}
We are certain that, in this rich research field, more promising solutions will be developed to address the above challenges, which will generate huge impacts on the evolution or even revolution of future communications.

\section{Acknowledgement}

We would like to acknowledge Professor Biing-Hwang Juang of Georgia Institute of Technology, Dr. Yiqun Ge and Mr. Rong Li of Huawei Technologies for their inspiring discussions and Dr. Zhijin Qin of Queen Mary University London for useful comments, which have helped improve the quality of the article.


\begin{thebibliography}{}

\bibitem{MIT1} N.~C~Thompson, K.~Greenewald, K. Lee, and G.~F.~Manso ``The computing limits of deep learning,'' {\em MIT Initiative on the Digital Economy Research Brief}, vol. 4,  2020.

\bibitem{Hao} H.~Ye, L.~Liang, G. Y. Li, and B.-H. F.~Juang, ``Deep learning-based end-to-end wireless communication systems with GAN as unknown channels,” {\em IEEE Trans. Wireless Commun.}, vol.~19, no.~5,~pp.~3133-3143, May 2020.

\bibitem{Q7} R.~K.~Srivastava, K.~Greff, and J.~Schmidhuber, ``Highway networks," at https://arXiv1505.00387v2, Nov.~2015.

\bibitem{Hopfield} R.~J.~McEliece, E.~C.~Posner E.~R.~Rodemich, and S.~S.~Venkatesh,``The capacity of the hopfield associative memory," {\em IEEE Trans. Inf. The.}, vol.~33, pp. 461-482, Jul. 1987.

\bibitem{Q9} H.-T.~He, S.~Jin, C.-K.~Wen, F.-F.~Gao, G.~Y.~Li, Z.~B.~Xu, ``Model-driven deep learning for physical layer communications," {\em IEEE Wireless Commun.}, vol.~26, no.~5, Oct.~2019.

\bibitem{Fred} B.-H. F. Juang, ``Accretionary learning with deep neural networks," personal communication.

\bibitem{Meta} T.~Hospedales, A.~Antoniou, P.~Micaelli, and A.~Storkey, ``Meta-learning in neural networks: a survey," {\em IEEE Trans. Pattern Analysis  Machine Intelligence}, May 2021.

\bibitem{Q3a} H.~B.~McMahan, E.~Moore, D.~Ramage, S.~Hampson, and B.~A.~Arcas, ``Communication-efficient learning of deep networks from decentralized data,'' in {\em Proc.~PMLR}, 2017.

\bibitem{Q3b} P.~Kairouz, et al., ``Advances and open problem in federated learning,'' {\em Foundations and Trends® in Machine Learning}, vol. 14, no. 1-2, pp. 1-210, Jun. 2021.

\bibitem{Q3Qin} Z.-J.~Qin, G.~Y.~Li, and H.~Ye, ``Federated learning and wireless communications,” to appear in {\em IEEE Wireless Commun.}, 2021, also at https://arxiv.org/abs/2005.05265.  


\bibitem{Hao2} H.~Ye, L.~Liang, and G.~Y.~Li, ``Decentralized learning with unreliable communications,” at https:// arxiv.org/abs/2108.02397, Jul. 2020.

\bibitem{Shannon} C.~E.~Shannon and W.~Weaver, {\em The Mathematical Theory of Communication}, The University of Illinois Press, 1949.

\bibitem{Semantic} J.~Bao, P.~Basu, M.~Dean, C.~Partridge, A.~Swami, W.~Leland, and J.~A.~Hendler, ``Towards a theory of semantic communication," {\em IEEE Network Science Workshop}, 2011, pp. 110-117.

\bibitem{Q5} N.~Farsad, M.~Rao, and A.~Goldsmith, ``Model-driven deep learning for physical layer communications," {\em IEEE ICASSP'18}, Apr. 2018.

\bibitem{Qin} H.~Q. Xie, Z.-J. Qin, G.~Y. Li, and B.-H. Juang, ``Deep learning enabled semantic communication systems," {\em IEEE Trans. Signal Process.}, vol. 69, pp. 2663-2675, May 2021.

\end{thebibliography}
\end{document}